# A PEP model of the electron

R.L. Collins, U.T Austin retired

**Abstract**

One of the more profound mysteries of physics is how nature ties together EM fields to form an electron. A way to do this is examined in this study. A bare magnetic dipole containing a flux quantum spins stably, and produces an inverse square **E= -vxB** electric field similar to what one finds from "charge". Gauss' law finds charge in this model, though there be none. For stability, a current loop about the waist of the magnetic dipole is needed and we must go beyond the classical Maxwell's equations to find it. A spinning **E** field is equivalent to an electric displacement current. The sideways motion of the spinning **E** (of constant magnitude) creates a little-recognized *transverse* electric displacement current about the waist. This differs from Maxwell's electric displacement current, in which **E** increases in magnitude. The sideways motion of **E** supports the dipolar **B** field, **B=vxE/c^2**. Beyond the very core of the magnetic dipole, each of these two velocities is essentially c and **vxE/c^2 = vx(-vxB)/c^2 = B**, the spinning **E** field wholly sourcing the dipolar **B** field. The anisotropy of the **vxB** field is cured by precession about an inclined axis. Choosing a Bohr magneton for the magnetic dipole and assuming it spins at the Compton frequency, Gauss' law finds Q = e. The **vxB** field, normally thought to be solenoidal, becomes instead a conservative field in this model. Charge is recognized as merely a mathematical construct, not fundamental but nevertheless useful. With charge deleted, and with addition of the transverse electric displacement current, Maxwell's equations can be written in terms of the **E** and **B** fields alone.

**Background**

In 1831, Faraday thought deeply on the nature of the Coulomb force and dismissed "force at a distance" in favor of an invisible static field he named the electric field **E**. He believed this **E** field to be sourced by charge, a concept that quantifies electrification. He did not concern himself with "what is charge", but merely accepted its existence. Charge has ever since been thought a real substance. Supposedly contained within or on the surface of a small spherical volume, charge is how we define the size of a charged elementary particle. But, what is charge? Charge has never been "explained". It just "is". There exists only one unit of charge, e, amongst all known free elementary charged particles.

While it is simple and convenient to attribute **E** to the presence of charge, several problems arise when we accept charge as a real substance. (a) Why does a particle not explode under Coulombic repulsion? (b) What is the size of a charged elementary particle? If the size of an electron were a measure of the volume containing the charge, then elastic Coulomb scattering should be able to find the size. Colliding electrons can be given enough energy to partially overcome Coulomb repulsion and cause an overlap of the spherical regions containing charge. This should diminish the Coulomb repulsion, alter the scattering pattern, and so let us measure size. It doesn't work. Experimentally, the electron always shows itself as a point particle without size. This is devastating, because the electric field increases without limit as r→0. The energy in the electric field is infinite for a point particle, and so an electron must have infinite mass. (c) The electron is known to carry angular momentum, but a point particle cannot. The problems associated with using charge as the source of the electric field continue to increase, the more we learn about electrons.

Pair production adds to these problems. If pair production somehow creates charge, charge can only be made from energetic EM fields. A few years ago, I noticed that a spinning dipolar **B** field creates, by **vxB**, an inverse square electric field that produces a net electric flux through a surrounding Gaussian surface. This implies charge creation. In effect, this was a discovery of the homopolar generator some 150 years after Faraday found it. A measurable d.c. voltage can be found by spinning a cylindrical bar magnet about its symmetry axis, with the leads of the voltmeter rubbing against the pole and the waist of the magnet. What is the seat of the emf? No one yet knows whether the dipolar **B** field from a bar magnet is itself spinning, or whether the voltage measured lies with metal cutting through a stationary **B** field. Faraday thought the **B** field stationary. His contemporary Weber thought the **B** field was spinning, which would create at every point in space a **vxB** electric field. This conundrum does not arise with the electron. An electron has a permanent magnetic moment and carries angular momentum, consistent with a spinning dipolar **B** field. This seemed promising for charge creation, but still lacked a mechanism that ties together EM fields stably in an elementary particle such as an electron. Herbert Jehle (1) again considered in the 1970's this **vxB** mechanism, attempting to model a PEP electron without postulating charge. He included magnetic flux quantization, but without a mechanism ensuring stability his work was not well received by the physics community.

Classical electromagnetic theory is ultimately based on the Lorentz force law that defines the **E** and **B** fields operationally, according to the force they exert on a test charge moving at velocity **v**.

**F = q(E + vxB)**  [1]

In the model discussed below, we will be dealing with moving (translating) **E** and **B** fields. The Lorentz force law assigns "**v**" to the velocity of the test charge. If, instead, the **B** field moves at **v**, a field very like **E** arises**, -vxB**. EM theory holds that charge is a fundamental though inexplicable attribute of some particles, and any suggestion that charge is not real must be disturbing.

## A PEP electron, from EM fields only

By PEP is meant purely electromagnetic particle. Several writers have considered the electron as an isolated spinning flux quantum, but were unable to explain its stability. (1,2) Consider the magnetic flux confined within a superconducting ring. One might expect that any magnetic flux whatsoever could be sustained, since any attempt to change the flux leads to an electric field along the ring. A superconductor cannot sustain an electric field, and so a correcting current immediately flows and denies any flux change. About 50 years ago, it was proposed (3) and experimentally verified (4) that magnetic flux is quantized. The experimental proof involves the step-wise increase of magnetic moment within a superconducting ring, when forced to accommodate larger and larger magnetic flux. Instead of London's predicted value of $\Phi_B = h/e$, it was found that the magnetic flux increased in units of h/2e. This was explained in terms of paired electrons, the current-carrying charge being 2e in a superconductor. The essence of the explanation for flux quantization is that the current-carrying charge exhibits a de Broglie wavelength that is related to the circumference of the ring. As with a Bohr atom, one needs a "stationary solution" for stability.

We have no reason to assume that that a flux quantum trapped within a superconducting loop is spinning, and it probably does not. Accepting this, one can transform to a rotating coordinate system so that the current-carrying charge 2e becomes stationary. Without current, what stabilizes the flux? The flux quantum is itself spinning, in this coordinate system, and spin is the only thing available to stabilize the flux. This study builds on this and finds that stability arises from a new kind of "displacement current", supplementing the one used by Maxwell to complete his four equations.

Consider the EM fields arising from a steady current "i" in a long straight conductor, shown in Figure 1. Maxwell's equations are:

$$\text{curl } \mathbf{E} = -\partial \mathbf{B}/\partial t \quad \text{curl } \mathbf{B} = \mu_0 i + \mu_0 \varepsilon_0 \partial \mathbf{E}/\partial t \quad \text{div } \mathbf{E} = \rho/\varepsilon_0 \quad \text{div } \mathbf{B} = 0 \qquad [2]$$

A circumferential **B** field is present outside the wire, perpendicular to the wire. Stoke's law applied to the curl **B** equation finds this to be of magnitude:

$$|\mathbf{B}| = \mu_0 i/2\pi r \qquad [3]$$

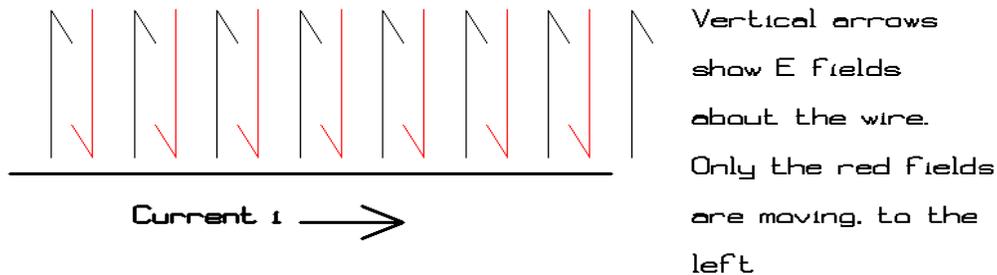

Figure 1. Electric fields outside a current-carrying wire.

The second of Maxwell's equations [2] tells us that current is the source of this **B** field. Another way to view the origin of this **B** field is to recognize that the negative current-carrying charges within the wire create a 1/r radial electrical field pointing inwards toward the wire (shown in red). This is everywhere matched by the fixed positive charges in the wire, which create a 1/r radial electric field (shown in black) pointing outwards. Although the net electric field outside the wire might be thought zero, the ingoing electric field is moving sideways relative to the outgoing electric field. A stationary test charge outside the wire finds no force, and [1] concludes there is no electric field. Not so! These fields are still there, but one of the fields is translating sideways. It follows that an electric field **E**, moving sideways, is wholly equivalent to a magnetic field, **B**.

$$\mathbf{B} = \mathbf{v} \times \mathbf{E}/c^2 \qquad [4]$$

It is then a matter of choice whether one describes the **B** field as arising from countermoving (sideways) **E** fields [4] or from current flow "i" [3]. It is not both. Attribution of **B** to current "i" is simpler, but the sideways motion of **E** is more tenable since it avoids the problem of "action at a distance." Action at a distance means that the establishment of a current instantly creates the **B** field everywhere in space. [4] makes it clear that the intermediary field is **E**, and we know that an **E** field cannot move faster than c. Although [4] would seem to have much to recommend it, many people have tried and failed to insert it into Maxwell's equations. Chubykalo (5,6) discusses these efforts at length. It may be that the poor acceptance of [4] has to do with the lack of a need for this enhancement of Maxwell's equations. In this model of a PEP electron, [4] is crucial and the model is not stable without it.

To find the Maxwell electric displacement current in curl **B** [2], we insert a capacitor as in Figure 2. With current "i" held constant, the **E** field within the capacitor increases with time. This formulation immediately avoids "action at a distance", since the changing **E** field between the plates of the capacitor cannot cause observable events instantly in all the space outside the capacitor. Maxwell realized that this increasing **E** field is equivalent to a current, and named it the electric displacement current $\varepsilon_0 \partial \mathbf{E}/\partial t$. This insight was used by Maxwell to complete the four famous equations known as Maxwell's laws. We will characterize this as *longitudinal*, to differentiate it from the other electric displacement current arising from the *transverse* motion of a field **E** whose magnitude is unchanging.

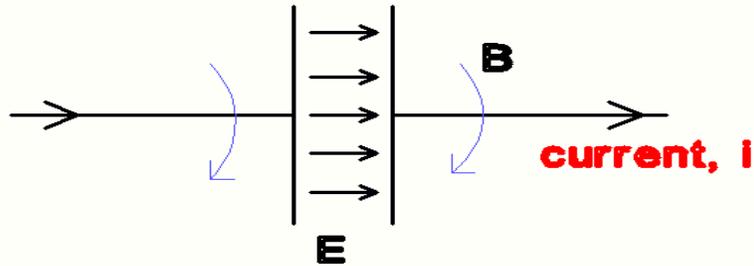

Figure 2. Geometry of Maxwell's displacement current.

With [4], there are now two electric displacement currents that source curl **B:** $\varepsilon_0 \partial \mathbf{E}/\partial t$ and $\varepsilon_0 \text{curl}(\mathbf{v} \times \mathbf{E})$. The current "i" in [2] is replaced by current density everywhere in space, $\varepsilon_0 \text{curl}(\mathbf{v} \times \mathbf{E})$, and one recognizes this as a *transverse* electric displacement current density. Conventional current "i" refers to the motion of charges, charges which we will find to be derivative rather than fundamental. A more fundamental form of the Maxwell equation for curl **B** may be had, in terms of EM fields only, upon replacing "i" with this *transverse* displacement current density $\varepsilon_0 \text{curl}(\mathbf{v} \times \mathbf{E})$. $\varepsilon_0 \mu_0 = 1/c^2$.

curl **B** = $\mu_0[\varepsilon_0 \partial \mathbf{E}/\partial t + \varepsilon_0 \text{curl}(\mathbf{v} \times \mathbf{E})]$     (longitudinal and transverse displacement currents)   [2']

Consider a magnetic dipole. A magnetic dipole of moment **μ** is conventionally sourced by a current loop about its waist. Without this current, classical EM theory holds that the **B** fields of the magnetic dipole will collapse and give rise to a surrounding toroidal Maxwellian displacement current, $\varepsilon_0 \partial \mathbf{E}/\partial t$. Not much help for a PEP model of an electron, since at best the dipole then oscillates between two configurations in which **μ** points in opposite directions. Since angular momentum is conserved, this means that the sign of the **-v×B** electric flux $\Phi_E$ reverses and Gauss' law finds no net charge has been produced. As noted above, spin seems to be necessary for stability. We next look to another reason why a spinning dipolar **B** field remains stable and does not oscillate.

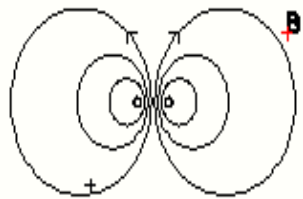

Figure 3. Side view of a magnetic dipolar **B** fields.

An inverse square **E= -vxB** field arises from the spinning **B** field in this model. Looking at the North end of the dipole, suppose that **B** spins clockwise. Does this **-vxB** field also spin, and if so in which direction and at what rate? Stability can be had by postulating that **-vxB** electric field also spins and in the same direction, giving rise to a transverse electric displacement current. Figure 4 depicts moving (rotating) lines of **E= -vxB** arising from current flow in a circular loop of conductor. The magnetic dipole here can be supported by either a real current in the loop or equivalently by a transverse "displacement current density" $\varepsilon_0$curl(**vxE**). In the vacuum of the PEP model, is this toroidal transverse displacement current sufficient to sustain the magnetic **B** field? For stability, we need that **vx(-vxB)**/$c^2$ = **B**. Happily, with each v=c, the transverse electric displacement current suffices.

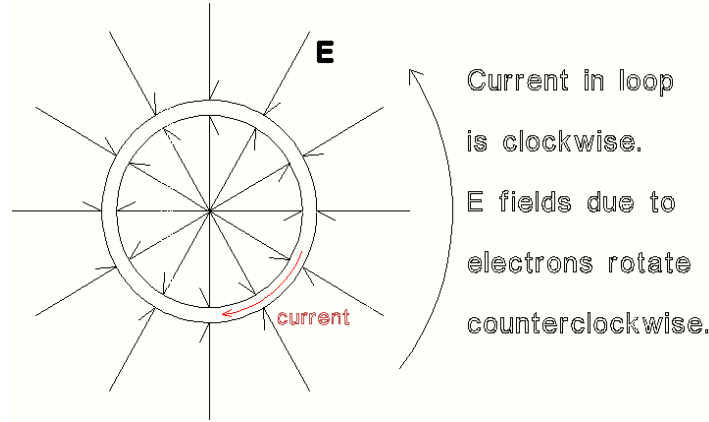

Figure 4. A current loop creates moving (rotating) **E** fields, viewed from south end of dipole.

This $\varepsilon_0$curl(**vxE**) is a different kind of electric displacement current density. It supplements Maxwell's longitudinal displacement current, in that here the **E** field moves sideways without change of magnitude. Not really that different, perhaps, because it is already implicit in Fig. 1 as an alternate to the current "i". This transverse displacement current is proportional to the transverse velocity, which in Figures 1 and 4 is limited by the Fermi velocity in the conductor. In the PEP model of the electron, absent a conductor, we may take it that this moves at v=c because **E** is not tied to moving electrons within a metal wire. Unconstrained **E** fields move in vacuum at c. Figure 4 is a bottom view of Figure 3, and shows the spinning **E** lines about the dipolar axis. The spinning magnetic field **B** creates the **E** field, and the **E** field also rotates at local velocity c and creates a transverse electric displacement current that supports the dipolar **B** field. Because the **E= -vxB** field is spinning so fast, each radial line depicting **E= -vxB** in this model should be drawn as a spiral. That is, each **E= -vxB** field line, as "r" increases, is not straight as suggested in Fig. 4. Instead, it lags behind at increasing "r" because each element moves sideways at c and the outer elements cannot keep up with the inner elements.

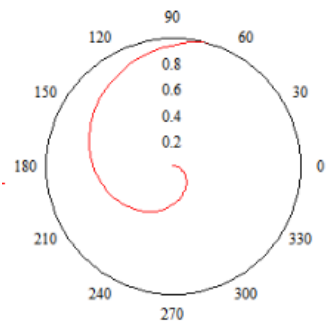

Figure 5. Typical line of spinning **vxB**, lagging with "r" because each segment of **vxB** moves at c.

The resulting model of a PEP electron is very simple, and this is fortunate since nature builds electrons so easily. A magnetic dipole containing a quantum of magnetic flux spins about its dipolar axis, creating an anisotropic inverse square **E**= -**vxB**, and is stabilized by a surrounding transverse electric displacement current loop. The isotropy one observes in the **E** field of an electron arises by time averaging of precession about an inclined axis. Ad hoc charge is not assumed, is not needed, and does not exist in this model. One recognizes that "charge" is not a real substance although it remains a useful mathematical construct. The model is unfortunately mute about what happens very close to center, where one cannot describe the **B** field by the far-field approximation shown below in [6].

## What is the "charge" on an electron?

We rely on Gauss' law to find the presence of charge. This model of an electron begins with a magnetic dipole containing a quantum of magnetic flux. Set spinning at angular velocity $\omega$, it creates an inverse square radial electric field -**vxB**. Gauss' law then finds whatever charge Q is enclosed within a spherical surface drawn about the model.

$$\Phi_E = \int (\mathbf{-vxB}) \cdot \mathbf{dS} = Q/\varepsilon_0 \qquad [5]$$

To calculate $\Phi_E$ for a spinning magnetic dipole having magnetic moment $\mu$, we need the tangential velocity of **B**, v= $\omega$r sin$\theta$, and the tangential component of **B**, i.e. $B_\theta$. In the far field,

$$\mathbf{B} = (\mu_0/4\pi)(\mu/r^3)[\mathbf{r_0}\, 2\cos\theta + \boldsymbol{\theta_0} \sin\theta] \quad \text{and so} \quad B_\theta = (\mu_0/4\pi)(\mu/r^3)\sin\theta \qquad [6]$$

$$vB_\theta = \omega r \sin\theta\, (\mu_0/4\pi)(\mu/r^3)\sin\theta \qquad [7]$$

This -**vxB** inverse square radial electric field [7] is created at every point in space by the spinning **B** field, and is anisotropic as shown by the $\sin^2\theta$ term. It remains inverse square at all "r", even as v $\to$ c, thanks to compensating SR factors. That is, the Lorentz transformation holds that **B** is $\gamma$ boosted and $\omega$ is diminished by 1/$\gamma$. Calculation of the charge Q is straightforward.

$$Q = \varepsilon_0 \Phi_E = \varepsilon_0 \int (\mu_0/4\pi)(\mu\omega/r^3)\sin\theta\, r \sin\theta\, 2\pi r^2 \sin\theta\, d\theta = 2\mu\omega/3c^2 \qquad [8]$$

Obvious first guesses for $\mu$ and $\omega$ are the Bohr magneton and the Compton frequency:

$$\mu = \mu_B = eh/4\pi m \qquad \text{and} \qquad \omega = 2\pi\, mc^2/h \qquad [9]$$

where m is the mass of an electron. [8] then finds Q=e/3. The anisotropy inherent in [7] can be cured by having the model spend 1/3 of the time with spin directed along each of 3 orthogonal directions. Most likely, this means that the spin precesses about another axis inclined at the magic angle arccos 1/$\sqrt{3}$ at a high rate. Our final choices of $\mu$ and $\omega$ must each be boosted by $\sqrt{3}$ because their experimentally measured (average) values are reduced by 1/$\sqrt{3}$ by this precession. With this, one finds

$$Q = e \qquad [10]$$

Many people have pondered over why we find in nature only one unit of charge. This model produces exactly the same unit of charge, e, no matter the mass m of the particle. The reason is that the product of $\mu\omega$ in [8] and [9] is independent of mass. Any elementary charged particle arising from this mechanism will have exactly the same unit of charge, e. The sign, of course, depends on the direction of spin. It was suggested above that **vx(-vxB)**/$c^2$ =**B**. With $\omega=2\pi mc^2/h$, the velocity **v" of** the spinning **B** field approaches c for r>$10^{-13}$ m. The **v** of the spinning **E= -vxB** is also postulated to be c, and in the same direction. The model is incomplete at very small distances, where [6] fails. Another problem is that the stability mechanism is wholly classical, much like the classical arguments for magnetic flux stabilization (but not quantization) within a superconducting ring. An as yet unknown quantization law selects for the mass of the electron.

## Conclusions

A "nude" magnetic flux quantum, configured as a magnetic dipole, will decay promptly unless supported by a toroidal current about the dipolar waist. A radial inverse square **E= -vxB** field, until now attributed to charge, arises from spin. Stability cannot be found within classical Maxwell's equations, which at best finds a toroidal electric displacement current $\varepsilon_0\partial\mathbf{E}/\partial t$ arising from the collapse of the **B** field. This predicts an oscillating magnetic dipole, which might be dynamically stable but which cannot create "charge". A little-known modification of Maxwell's equations provides stability. The sources of **B** are, conventionally, Maxwell's displacement current $\varepsilon_0\partial\mathbf{E}/\partial t$ and/or an actual current, i. The needed modification is found upon recognizing that this "i" is equivalent to a current density $\varepsilon_0$curl(**vxE),** in which an **E** field translates without change of magnitude. A spinning **E** field constitutes a *transverse* electric displacement current that supports the magnetic flux quantum and prevents its decay. Assuming the time-averaged magnetic moment $\mu$ to be a Bohr magneton, and with an averaged spin rate equal to the Compton frequency, Gauss' law tells us this model contains "charge" Q = e, even though the model contains no charge.

Is this merely taxonomy, a renaming of charge in favor of rotating **B** fields as the source of a static electric field? Not so, for several reasons. The **vxB** electric field in this model is not static, but undulates at a high rate because of precession. Maxwell's equations assume the existence of a real substance we call charge, sourcing a static electric field **E**. This study concludes that charge is not a real substance. If we want Maxwell's equations in fundamental form, they must deal with fields alone and not with "charge". We can retain the electric field **E** in Maxwell's equations, realizing that its only source is a changing **B** field. Likewise, **B** derives solely from a changing **E** field. This electric field **E= -vxB** can be either solenoidal or conservative, according to the geometry by which it is produced. This is discussed in the appendix. Since charge is not a real substance, size of a charged elementary particle cannot be defined by whatever volume contains the charge. Measured by its EM fields, every electron is infinitely large. Measured by Coulomb repulsion, an electron is essentially a point particle. At very small radial distances, the dipolar formula [6] fails and one must then take into account the size and structure of the magnetic dipole. This is beyond the scope of this study.

The current paradigm in particle physics does not seek to understand structure in electrons at the PEP level. Instead, it accepts that these particles are always popping up out of the vacuum sea and as quickly disappear unless given sufficient energy. Since electrons are "fundamental particles", one is unconcerned that such particles carry "charge". Questions about the structure and composition of an electron are meaningless and scattering experiments that cannot define "size" are no cause for worry.

Most advances in physics are "discoveries". A few are "undiscoveries", as with the luminiferous ether. Perhaps this undiscovery of the substance called charge, and of a static electric field **E** arising from charge, will also be considered advances. I find the PEP concept compelling, in which particles are ultimately composed solely of EM fields. Any model of an elementary charged particle must be simple, since nature builds them so easily. We may have finally reached the root of all things, the magnetic field **B,** although the reasoning remains somewhat circular as to whether **E** creates **B** or vice versa. I prefer **B** as fundamental, if only because magnetic flux is known to be quantized. This model finds "charge" of an elementary particle to be independent of mass, which helps explain why we find only the one unit of charge.

This study denies the concept of charge as a real substance, and finds that **E** and **B** are sourced solely by changes in the other field. The changes of the "other" field include (a) its magnitude and/or (b) its transverse velocity. Maxwell's equations recognize that **E** is sourced by the changing magnitude of **B** (longitudinal) and the Lorentz force law (transverse) recognizes that -**vxB** is very like **E**. The sources of **B** are treated differently. Its sources are (a) a changing magnitude of **E** and/or (b) an actual current, "i". This study suggests that "i" be replaced with the *transverse* electric displacement current density, $\varepsilon_0$curl(**vxE**). This is more general, since it includes the effects of a real current and includes the effects of a spinning **E** field that can exist in the absence of a real current. It also avoids "action at a distance". With this, **vxE** is the transverse source and $\partial$**E**/$\partial$t is the longitudinal source of **B**. Likewise, -**vxB** is the transverse source and -$\partial$**B**/$\partial$t is the longitudinal source of **E**. It follows that there cannot exist a free-standing **E** or **B** field, separate from the other field. Whenever one finds an **E** field, one must find its source as a changing **B** field. Where one finds a **B** field, one must find its source as a changing **E** field. In the model of an electron, a stable isolated dipolar magnetic field must be sourced. The source could be either (a) a surrounding *longitudinal* Maxwellian electric displacement current in which an **E** field increases in magnitude at the expense of a shrinking **B** field or (b) a surrounding transverse electric displacement current in which **E** spins but does not change magnitude and in which case the **B** field is stable and does not shrink. In the event, stability arises from the latter. Einstein's SR recognizes that **E** and **B** are somewhat interchangeable, according to motion (transverse) relative to the observer. This conforms with a velocity change in the transverse (but not the longitudinal) sources discussed above. GR includes the possibility of acceleration, and this implies that longitudinal effects must then also be considered. The modified Maxwellian equations become fully symmetric.

## Appendix: Electromagnetism, at the level of E and B fields.

Classical electromagnetic theory attempts to explain the relationships between charge and electric and magnetic fields. These concepts arose separately, and through the genius of Faraday and Maxwell were combined and interrelated. We have also learned, from Einstein's insight, that **E** and **B** are interchangeable according to the observer's motion and have no independent existence. In a vacuum, the Lorentz force exerted on a test charge q defines operationally the **E** and **B** fields.

**F** = q(**E**+**vxB**) [1A]

About 150 years ago, Maxwell collected the known EM relationships, added the electric displacement current $\varepsilon_0\partial$**E**/$\partial$t to curl **B**, and found the almost symmetric equations known and revered as Maxwell's equations. In a vacuum, with $\rho$ being charge density,

$$\text{curl } \mathbf{E} = -\partial \mathbf{B}/\partial t \quad \text{curl } \mathbf{B} = \varepsilon_0 \mu_0 \, \partial \mathbf{E}/\partial t + \mu_0 \mathbf{i} \quad \text{div } \mathbf{E} = \rho/\varepsilon_0 \quad \text{div } \mathbf{B} = 0 \quad [2A]$$

Maxwell's equations describe two very different kinds of electric field, **E**. In the first equation, **E** arises from a changing **B** field and is solenoidal. Solenoidal means that the field has neither sources nor sinks, and so the "lines of force" used to visualize the field are continuous loops. The third equation describes **E** arising from charges, and this field is conservative. Among other things, conservative means that the lines representing **E** are sourced or sinked by charge. The electric force at charge q due to another charge Q, with **E** as intermediary, is described by Coulomb's law:

$$F = (1/4\pi\varepsilon_0)qQ/r^2 \qquad E = (1/4\pi\varepsilon_0)Q/r^2 \qquad \mathbf{F} = q\mathbf{E} \qquad [3A]$$

The **E** field concept was introduced by Faraday. Recognizing that "action at a distance" is not logical, he introduced the electric field **E** as the tangible connection between charges. When sourced by charge, **E** is a conservative field. Conservative means that the work needed to move a charge q from one point to another is independent of the path. This differs from the solenoidal electric field derived from a changing **B** field, in which a test charge gains (or loses, depending on the direction) energy each time it transits a loop of the **E** field. It seemed plausible to Faraday, and has to us ever since, that the conservative **E** field is static, real, carries energy, and arises solely from charge. This differs from the **-vxB** electric field from a PEP electron discussed in this paper, a conservative field that undulates at the precession frequency and only seems static because current measuring instruments are unresponsive to rapid oscillation at the Compton frequency. Maxwell's equations tell us that **E**= **-vxB** is always solenoidal, because no charge is involved, and that **E** arising from charge is always conservative.

These two kinds of electric field, **E** and **-vxB**, are so different that it is surprising that we even continue to use the same symbol, **E**. Lorentz' force law recognizes both, one described as **E** (conservative, from charges) and the other as **vxB** (solenoidal). Until now, it has not been thought that anything other than "charge" could produce a conservative inverse square electric force upon a test charge. It will be shown next that **-vxB** in this model can do just that.

Lets look into a mechanism by which the **-vxB** field, itself intrinsically solenoidal and hence divergenceless, can source or sink **-vxB** and so become a conservative electric field. Consider a skinny single loop of **B** field enclosing the blue circle in figure 1A. Not to worry about how to produce it, we just assume it and it is so by construction (thanks a lot, theorists). The **B** field orientation is counterclockwise. When set spinning about a vertical diameter, what do we find? On the left side, **-vxB** is directed towards the center and on the right side, **-vxB** is directed away from center. What do we call an entity that produces such an electric field? A spinning *electric dipole*, with no net charge. Now move the vertical spin axis to the left, tangent to the loop. Then spin it again and consider the **-vxB** field. On the left, it vanishes and on the right it is directed outwards and with twice the magnitude. In the first case, a Gaussian surface drawn about the figure finds zero net electric flux and hence no charge. The surface integral of **-vxB** = 0. In the second case, the Gaussian surface finds a positive definite electric flux and so concludes that charge is present within. What do we call the entity that produces such a field? Gauss' law calls it "*charge*". This mechanism, without reliance on any ad hoc charge, can source its own **-vxB** electric field and also sink electric fields produced elsewhere. What, then, is charge? Charge is a measure of electrification, assumed incorrectly by Faraday to be a real substance that "sources" a static inverse square electrical field **E**. This simplified loop of **B** can readily be expanded to form a symmetric magnetic dipole, as in the proposed PEP electron model. The distinction between a conservative and a solenoidal electrical field is not so great as has been generally accepted, always arising from **-vxB** but converting from one to the other with slight changes of geometry within the mechanism that produces the electric field from moving **B** fields.

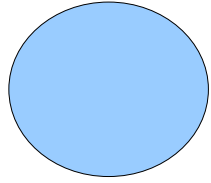

Figure 1A. Counterclockwise loop of B field, initially spinning about a vertical diameter.

It follows that Maxwell's "curl" equations can be written in a more symmetrical and fundamental form using EM fields only, excluding any reference to "charge":

curl **B** = $\mu_0\varepsilon_0$ [$\partial$**E**/$\partial$t + curl(**vxE**)]     instead of     curl **B** = $\mu_0\varepsilon_0$ $\partial$**E**/$\partial$t + $\mu_0$i     [4A] curl **E** = $-\partial$**B**/$\partial$t + curl(**-vxB**)     instead of     curl **E** = $-\partial$**B**/$\partial$t     [5A]

$1/c^2 = \mu_0\varepsilon_0$. [4A] recognizes the fundamental sources of **B** as two *electric* displacement currents, one being the changing magnitude of **E** (longitudinal) and the other a rotating or translating **E** field (transverse). Rotation about a point far away is equivalent to translation. This both simplifies and makes more symmetric Maxwell's laws. It follows that **B** arises from changes in **E**, and **E** arises from changes in **B**. And, from nothing else! Charge and current are useful derivative constructs, but have no place in any really fundamental form of Maxwell's equations.

It may be useful to write Maxwell's 4 equations as 6 equations, in even simpler form.

div **E** = 0                        div **B** = 0                        [6A]

That is, fundamentally, there are no electric or magnetic monopoles. Changes in one field are the only source of the other. These changes involve magnitude and/or translation. When change is confined to magnitude, we have the longitudinal "curl" equations:

curl **B** = $\mu_0\varepsilon_0$ $\partial$**E**/$\partial$t = ($\partial$**E**/$\partial$t)/$c^2$  (**E** increasing)     curl **E** = $-\partial$**B**/$\partial$t  (**B** increasing)     [7A]

When the changes involve translation, the curl operator is not needed. The transverse equations that complete Maxwell's equations are then, with **c** for the velocities of **E** and **B**:

**B** = $\mu_0\varepsilon_0$(**vxE**) = **vxE**/$c^2$   (**E** spinning)                    **E** = -**vxB**   (**B** spinning)         [8A]

Using the "bac cab" rule of vector multiplication, **Ax(BxC)= B**(A.C)-**C**(A.B), one easily finds, with v=c, that the dipolar magnetic field **B** in this model is supported fully by the spinning **E** field.

**B** = **vx(-vxB)**/$c^2$ = [**v**(**v.B**) - **B**(-**v.v**)]/$c^2$ = **B**     [9A]

Notice that **v.B**=0 because **v** is perpendicular to the dipolar **B** field and **v.v**= $c^2$ because these velocities lie in the same direction.

The only sources of **E** and **B** fields are changes in the other field. The factor $1/c^2$ is needed for dimensional purposes. The discussion above, especially that dealing with Figure 1A, shows how one can tie these fields into a configuration that resembles an electric monopole. Although this electric monopole sources a conservative inverse square electric field, it does so without inclusion of or need for the substance we have called "charge". If these arguments are persuasive, the basic laws of electromagnetism need to be modified. The static conservative Coulombic **E** field, arising from charge, does not exist. The **E** field about a charged particle undulates at a high rate, and only seems static because our measuring instruments cannot respond so quickly. Charge is not a real substance, and only serves as a mathematical convenience when dealing with the electrical forces exerted by one particle on another. A compelling reason for "why no magnetic monopoles?" is now clear. So long as there were thought to exist fundamental electrical monopoles, a deep belief in symmetry demanded that we consider and look for magnetic monopoles. This study concludes that there are, fundamentally, no electrical monopoles either. And so, symmetry has been recovered.